\documentclass[twoside,11pt]{article}

\usepackage[preprint]{style}

\usepackage{mathptmx}
\usepackage{amsmath}
\usepackage{graphicx}
\usepackage{subfig}
\usepackage{color}
\usepackage{xspace}
\usepackage{mathtools}
\usepackage{booktabs}
\usepackage{enumitem}
\usepackage[T1]{fontenc}

\usepackage{listings}
\usepackage{color}

\usepackage{lastpage}

\definecolor{dkgreen}{rgb}{0,0.6,0}
\definecolor{gray}{rgb}{0.5,0.5,0.5}
\definecolor{mauve}{rgb}{0.58,0,0.82}

\newcommand{\dowhy}{\texttt{DoWhy}\xspace}
\newcommand{\dowhygcm}{\texttt{DoWhy-GCM}\xspace}

\jmlrheading{25}{2024}{1-\pageref{LastPage}}{11/22; Revised 12/23}{5/24}{22-1258}{Patrick Blöbaum, Peter Götz, Kailash Budhathoki, Atalanti A. Mastakouri and Dominik Janzing}

\ShortHeadings{\dowhygcm: An Extension of \dowhy for Causal Inference in Graphical Causal Models}{Blöbaum, Götz, Budhathoki, Mastakouri, Janzing}
\firstpageno{1}

\begin{document}
\title{\dowhygcm: An Extension of \dowhy for Causal Inference in Graphical Causal Models}

\author{%
\name Patrick Blöbaum \email bloebp@amazon.com
\AND
\name Peter Götz \email pego@amazon.com
\AND
\name Kailash Budhathoki \email kaibud@amazon.com
\AND
\name Atalanti A. Mastakouri \email atalanti@amazon.com
\AND
\name Dominik Janzing \email janzind@amazon.com\\
\addr{Amazon Web Services}
}

\editor{\vspace*{-0.3cm}Joaquin Vanschoren}

\maketitle

\begin{abstract}%   <- trailing '%' for backward compatibility of .sty file
We present \dowhygcm, an extension of the \dowhy Python library, which leverages graphical causal models. Unlike existing causality libraries, which mainly focus on effect estimation, \dowhygcm addresses diverse causal queries, such as identifying the root causes of outliers and distributional changes, attributing causal influences to the data generating process of each node, or diagnosis of causal structures.
With \dowhygcm, users typically specify cause-effect relations via a causal graph, fit causal mechanisms, and pose causal queries---all with just a few lines of code. The general documentation is available at \href{https://www.pywhy.org/dowhy}{https://www.pywhy.org/dowhy} and the \dowhygcm specific code at \href{https://github.com/py-why/dowhy/tree/main/dowhy/gcm}{https://github.com/py-why/dowhy/tree/main/dowhy/gcm}.
\end{abstract}


\section{Introduction}
Analyzing complex systems beyond mere correlations is becoming increasingly important in modern data science~\citep{causality_book,Pearl2009}. In order to analyze a system and to understand the interaction between components, modeling their causal interactions is crucial. One of the wide-spread applications of causality is in \textit{effect estimation} problems, where the goal is to estimate the effect of an intervention on one variable on another~\citep{Imbens2015,Rubin1974}. The Python library \dowhy~\citep{dowhy} initially only focused on addressing such type of problems, but causal questions often span beyond effect estimation problems~\citep{causality_book,pmlr-v238-janzing24a,rca_outliers,distribution_change,janzing2020feature}. 

\begin{figure}[tb]
    \centering
    \vspace*{-0.9cm}
    \includegraphics[scale=0.38,keepaspectratio]{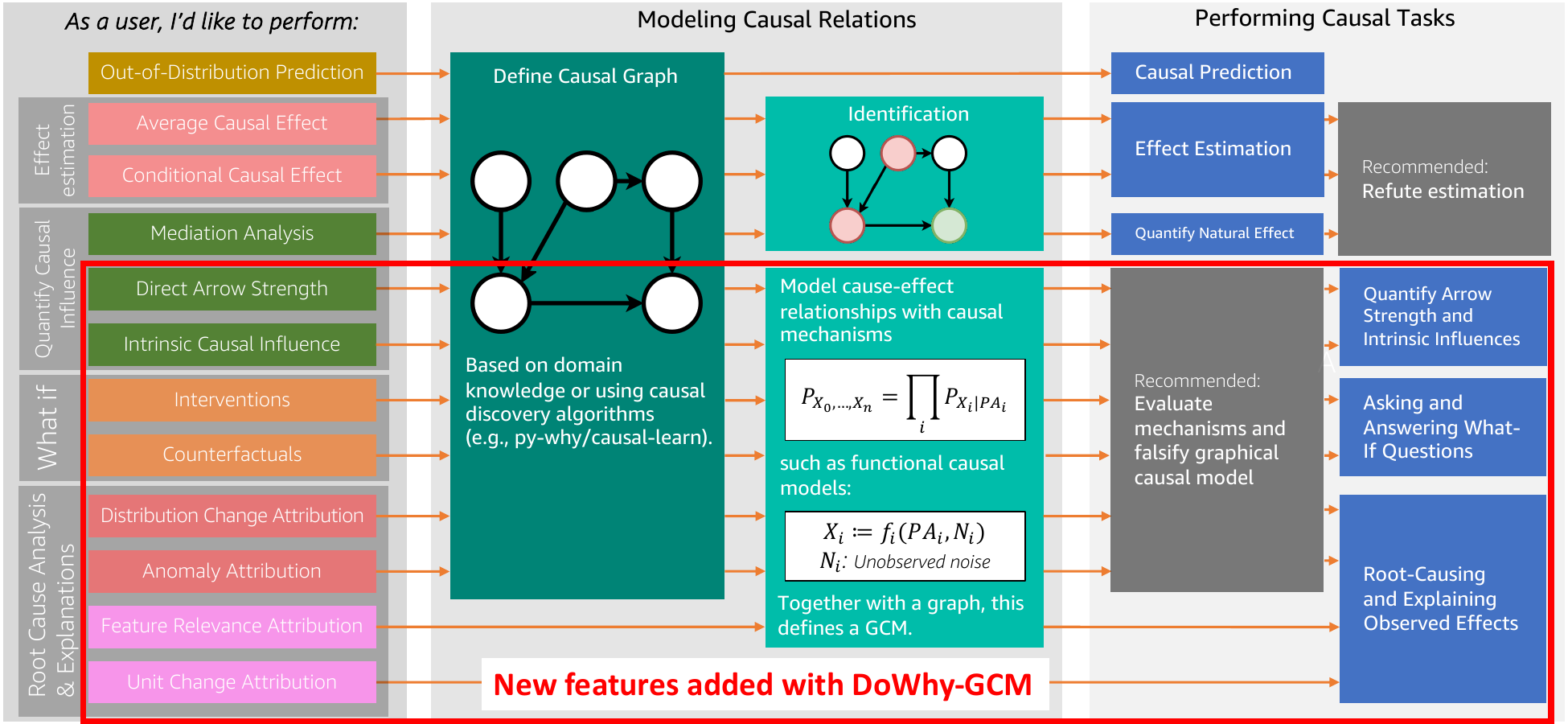}
    \caption{\dowhygcm complements \dowhy by allowing to address a wide range of causal questions by utilizing graphical causal models. The graphical structure is a common data type across most of the features, which ensures interoperability.}
    \label{fig:features}%
\end{figure}

\dowhygcm is a complementary extension of \dowhy and leverages graphical causal models (GCM) \citep{Pearl2009} to answer a broader range of causal questions (see Figure \ref{fig:features} for an overview). To get started with \dowhygcm, users provide two objects: a directed acyclic graph (DAG) representing the causal relationships between variables in a system, and \textit{tabular observational data} corresponding to the variables in the DAG, which can be continuous, discrete or categorical. They then fit a GCM, and ask causal questions, such as identifying root causes of outliers~\citep{rca_outliers}, quantitatively attribute distributional changes to the data generating mechanisms in place at each node~\citep{distribution_change}, quantify strength of causal influences~\citep{causalstrength,pmlr-v238-janzing24a,janzing2020feature}, or point- and population-level based estimation of counterfactuals~\citep{Pearl2009,causality_book} (see Section \ref{sec:func} for more).

\dowhygcm has a modular design to allow easy-integration with third-party libraries. To this end, we define interfaces for key components of the library, such as causal graph, causal model, and causal mechanism. This allows users to contribute their own algorithms and models.
For example, if a user knows the functional relationship between a variable and its parents, they can assign a custom implementation of the causal mechanism to that variable. In addition, for several causal queries, we provide wrappers on top of popular existing third-party libraries, such as \texttt{scikit-learn}~\citep{scikit-learn} and \texttt{SciPy}~\citep{2020SciPy-NMeth} and, thus, support wide range of parametric as well as non-parametric models. While \dowhygcm offers a general framework for modeling a GCM, it comes with various native implementations of certain algorithms (see Section \ref{sec:func}).

\section{\dowhygcm Three-Step Recipe}
\dowhygcm implements a general purpose framework for modeling graphical causal models, which is based on the formal framework developed by Judea Pearl~\citep{Pearl2009}. Cause and effect relationships are represented via directed edges and the data generation process of each node in the graph given its parents is described by a so-called \textit{causal mechanism}. These mechanisms are assumed to be modular, i.e., we can independently change the causal mechanism of one variable without affecting the causal mechanisms of other variables in the system (\citet[Ch.~2]{causality_book}). Typically answering a causal question using GCMs involves 3 key steps, which \dowhygcm embraces.

\textbf{1/ Model cause-effect relationships in a graphical causal model:} The first step is to model the cause-effect relationships between variables through a causal graph using any supporting graph library that abides by our interface, e.g. \texttt{NetworkX}~\citep{SciPyProceedings_11}. In addition, we assign a causal mechanism at each node in the causal graph that allows us to model the data generation process. A causal mechanism of a node yields its conditional distribution given its parents and modeling theses mechanisms explicitly at each node allows us to answer a wide range of causal questions beyond typical effect estimation tasks. As a root node does not have any parents, we obtain its distribution from either the empirical distribution or a parametric model. We refer to the causal graph and its causal mechanisms as a GCM.

Modeling the generating process of each node explicitly, however, requires additional assumptions about how the causal mechanisms look like. While the user has the flexibility to define their own custom mechanisms, \dowhygcm also provides an API to automatically infer the causal mechanisms from observational data. For this, \dowhygcm supports additive noise model (ANM) \citep{Hoyer} and post non-linear models \citep{Zhang_UAI} out of the box, which come with certain modeling assumptions but are flexible and restrictive enough to render tasks like estimation of counterfactuals solvable. To further illustrate this, let us use a simple causal graph $X \to Y$, where we model the marginal distribution $P_X$ and the conditional distribution $P_{Y \mid X}$. Using, e.g., an ANM here for $Y$, we obtain $X \coloneqq N_X$, and $Y \coloneqq f(X) + N_Y$, where $N_X$ and $N_Y$ are assumed to be independent. In \dowhygcm, noises $N_X$ and $N_Y$ can be represented by a probability distribution, whereas the function $f$ can be any prediction function (e.g., (non-)linear regression). If the causal mechanism permits, \dowhygcm can even reconstruct the noise based on a given observation, which is a requirement for estimating counterfactuals in Pearl's framework.

\textbf{2/ Fit the causal mechanisms:} After assigning causal mechanisms to nodes in the causal graph, the next step is to learn the parameters of those mechanisms from data in case of parametric models. Note that \dowhygcm also allows users to provide their own "ground truth" causal mechanism. After this step, the GCM is ready for causal reasoning tasks.

\textbf{3/ Ask a causal question:} With a fitted GCM, users can then answer a wide range of causal questions, where the same fitted GCM can be reused for different queries.

All these step above can be done with a few lines of code:
\begin{lstlisting}
# Define structural causal model and automatically assign models
causal_model = gcm.StructuralCausalModel(nx.Digraph([("X", "Y"), ("Y", "Z")]))
gcm.auto.assign_causal_mechanisms(causal_model, samples_df)

# Fit generative models
gcm.fit(causal_model, samples_df)

# Optional: Evaluate causal model
gcm.evaluate_causal_model(causal_model, samples_df)

# Causal model is now ready for different causal queries
gcm.<causal query>(causal_model, ...)
\end{lstlisting}

\section{\dowhygcm Features: $\texttt{gcm}.\langle \texttt{causal query} \rangle$ }
\label{sec:func}
This section covers some exemplary causal queries users can ask with a fitted GCM object:
\begin{description}[noitemsep, leftmargin=*]
    \setlength\itemsep{0em}
    \item[Causal Model Evaluation:] These methods check statistically falsifiable assumptions in a causal model, e.g., reject a graph or evaluate assigned causal mechanisms given data.
    \begin{itemize}[leftmargin=*,label=>>]
        \setlength\itemsep{0em}
        \item \href{https://github.com/py-why/dowhy/blob/e7bd2ad84fbab4f2ddab2c4df2ea70178080aab7/dowhy/gcm/falsify.py#L510}{\texttt{gcm.falsify\_graph}}: Falsify a given DAG using observational data~\citep{eulig2023falsifying}.
        \item \href{https://github.com/py-why/dowhy/blob/e7bd2ad84fbab4f2ddab2c4df2ea70178080aab7/dowhy/gcm/model_evaluation.py#L303}{\texttt{gcm.evaluate\_causal\_model}}: Evaluate causal mechanisms and modeling assumptions using observational data.
    \end{itemize}
    \item[Quantify Causal Influence:] These quantify causal influences, such as the strength of an edge or the intrinsic influence of nodes on a target.
    \begin{itemize}[leftmargin=*,label=>>]
        \setlength\itemsep{0em}
        \item \href{https://github.com/py-why/dowhy/blob/e7bd2ad84fbab4f2ddab2c4df2ea70178080aab7/dowhy/gcm/influence.py#L34}{\texttt{gcm.arrow\_strength}}: Quantifies the strength of an edge~\citep{causalstrength}.
        \item \href{https://github.com/py-why/dowhy/blob/e7bd2ad84fbab4f2ddab2c4df2ea70178080aab7/dowhy/gcm/influence.py#L216}{\texttt{gcm.intrinsic\_causal\_influence}}: Quantifies contribution of a source node to the uncertainty in a target node~\citep{pmlr-v238-janzing24a}.
    \end{itemize}
    \item[What-If:] These methods estimate the effect of interventions which can be more general beyond standard atomic interventions, and compute counterfactuals~\citep{causality_book}.
    \begin{itemize}[leftmargin=*,label=>>]
        \setlength\itemsep{0em}
        \item \href{https://github.com/py-why/dowhy/blob/e7bd2ad84fbab4f2ddab2c4df2ea70178080aab7/dowhy/gcm/whatif.py#L27}{\texttt{gcm.interventional\_samples}}: Performs intervention on nodes in the causal graph.
        \item \href{https://github.com/py-why/dowhy/blob/e7bd2ad84fbab4f2ddab2c4df2ea70178080aab7/dowhy/gcm/whatif.py#L105}{\texttt{gcm.counterfactual\_samples}}: Estimates counterfactual data for observed data if we were to perform specified interventions.
    \end{itemize}
    \item[Attribution:] These methods attribute observed effects (e.g., outliers, distributional changes) to their root causes (i.e., nodes upstream in the causal graph).
    \begin{itemize}[leftmargin=*,label=>>]
        \setlength\itemsep{0em}
        \item \href{https://github.com/py-why/dowhy/blob/e7bd2ad84fbab4f2ddab2c4df2ea70178080aab7/dowhy/gcm/anomaly.py#L90}{\texttt{gcm.attribute\_anomalies}}: Quantifies contributions to anomalies~\citep{rca_outliers}.
        \item \href{https://github.com/py-why/dowhy/blob/e7bd2ad84fbab4f2ddab2c4df2ea70178080aab7/dowhy/gcm/distribution_change.py#L88}{\texttt{gcm.distribution\_change}}: Quantifies the contribution of each node to the change in the distribution of the target~\citep{distribution_change}.
        \item \href{https://github.com/py-why/dowhy/blob/e7bd2ad84fbab4f2ddab2c4df2ea70178080aab7/dowhy/gcm/feature_relevance.py#L19}{\texttt{gcm.parent\_relevance}}: Quantifies the relevance of a causal parent with respect to the causal mechanism by explicitly incorporating the noise~\citep{Lundberg2017,janzing2020feature}.
        \item \href{https://github.com/py-why/dowhy/blob/e7bd2ad84fbab4f2ddab2c4df2ea70178080aab7/dowhy/gcm/unit_change.py#L34}{\texttt{gcm.unit\_change}}: Quantifies the contributions of input features and prediction mechanisms to the change in the value of the target for a statistical unit~\citep{budhathoki2022explaining}.
    \end{itemize}
\end{description}

\section{Code design}
\dowhygcm follows functional programming principles to focus on causal questions and avoid issues like code delegation, duplication, bookkeeping, statefulness, and scalability problems with ever-growing class APIs. It leverages existing, commonly used libraries like \texttt{NetworkX}~\citep{SciPyProceedings_11}, \texttt{NumPy}~\citep{harris2020array}, and \texttt{Pandas}~\citep{mckinney-proc-scipy-2010} to avoid reinventing the wheel. The API is designed to be functional - operating on a GCM object and returning results to prevent statefulness. It provides sensible defaults for methods, offers various convenience functions to simplify usage while allowing full customization, and enables inspection by exposing the various GCM components and models explicitly.

\section{Discussion}
We introduced \dowhygcm, an extension of \dowhy, that complements \dowhy's existing effect estimation features by adding features for graphical causal models operating on the same causal graphs. Graphical causal models allow us to look at a system as a whole, not just the effect of one variable on another, allowing us to address causal questions beyond effect estimation. That is, we model cause-effect relationships between all variables explicitly with causal mechanisms representing probabilistic models and fit them based on observational tabular data. The API is designed with a focus on modularity, allowing easy integration of custom models, algorithms, or other third-party libraries. The scalability of the provided algorithms heavily depend on the inference complexity of the used models, the number of variables in the causal graph, the sample size, or the structure of the causal graph. In contrast to other Python causality libraries such as \texttt{EconML}~\citep{econml}, \texttt{CausalML}~\citep{chen2020causalml}, \texttt{cdt}~\citep{cdt}, \texttt{DiCE}~\citep{mothilal2020dice}, \texttt{Tetrad}~\citep{ramsey2018tetrad}, \texttt{CausalNex}~\citep{Beaumont_CausalNex_2021} or \texttt{WhyNot}~\citep{miller2020whynot} that primarily focus on effect estimation, \dowhygcm offers features and novel algorithms that explicitly leverage graphical causal models.


\newpage

\begin{thebibliography}{27}
\providecommand{\natexlab}[1]{#1}
\providecommand{\url}[1]{\texttt{#1}}
\expandafter\ifx\csname urlstyle\endcsname\relax
  \providecommand{\doi}[1]{doi: #1}\else
  \providecommand{\doi}{doi: \begingroup \urlstyle{rm}\Url}\fi

\bibitem[Battocchi et~al.(2019)Battocchi, Dillon, Hei, Lewis, Oka, Oprescu, and
  Syrgkanis]{econml}
K.~Battocchi, E.~Dillon, M.~Hei, G.~Lewis, P.~Oka, M.~Oprescu, and
  V.~Syrgkanis.
\newblock {EconML}: {A Python Package for ML-Based Heterogeneous Treatment
  Effects Estimation}.
\newblock https://github.com/microsoft/EconML, 2019.
\newblock Version 0.x.

\bibitem[Beaumont et~al.(2021)Beaumont, Horsburgh, Pilgerstorfer, Droth,
  Oentaryo, Ler, Nguyen, Ferreira, Patel, and Leong]{Beaumont_CausalNex_2021}
P.~Beaumont, B.~Horsburgh, P.~Pilgerstorfer, A.~Droth, R.~Oentaryo, S.~Ler,
  H.~Nguyen, G.~A Ferreira, Z.~Patel, and W.~Leong.
\newblock {CausalNex}, 10 2021.
\newblock URL \url{https://github.com/quantumblacklabs/causalnex}.

\bibitem[Budhathoki et~al.(2021)Budhathoki, Janzing, Bloebaum, and
  Ng]{distribution_change}
K.~Budhathoki, D.~Janzing, P.~Bloebaum, and H.~Ng.
\newblock Why did the distribution change?
\newblock In Arindam Banerjee and Kenji Fukumizu, editors, \emph{Proceedings of
  The 24th International Conference on Artificial Intelligence and Statistics},
  volume 130 of \emph{Proceedings of Machine Learning Research}, pages
  1666--1674. PMLR, 13--15 Apr 2021.
\newblock URL \url{https://proceedings.mlr.press/v130/budhathoki21a.html}.

\bibitem[Budhathoki et~al.(2022{\natexlab{a}})Budhathoki, Michailidis, and
  Janzing]{budhathoki2022explaining}
K.~Budhathoki, G.~Michailidis, and D.~Janzing.
\newblock Explaining the root causes of unit-level changes, 2022{\natexlab{a}}.
\newblock URL \url{https://arxiv.org/abs/2206.12986}.

\bibitem[Budhathoki et~al.(2022{\natexlab{b}})Budhathoki, Minorics,
  Bl{\"o}baum, and Janzing]{rca_outliers}
K.~Budhathoki, L.~Minorics, P.~Bl{\"o}baum, and D.~Janzing.
\newblock Causal structure-based root cause analysis of outliers.
\newblock In \emph{International Conference on Machine Learning}, pages
  2357--2369. PMLR, 2022{\natexlab{b}}.

\bibitem[Chen et~al.(2020)Chen, Harinen, Lee, Yung, and Zhao]{chen2020causalml}
H.~Chen, T.~Harinen, J.~Lee, M.~Yung, and Z.~Zhao.
\newblock {CausalML}: Python package for causal machine learning, 2020.
\newblock URL \url{https://arxiv.org/abs/2002.11631}.

\bibitem[Eulig et~al.(2023)Eulig, Mastakouri, Blöbaum, Hardt, and
  Janzing]{eulig2023falsifying}
E.~Eulig, A.~A. Mastakouri, P.~Blöbaum, M.~Hardt, and D.~Janzing.
\newblock Toward falsifying causal graphs using a permutation-based test, 2023.
\newblock URL \url{https://arxiv.org/abs/2305.09565}.

\bibitem[Hagberg et~al.(2008)Hagberg, Schult, and Swart]{SciPyProceedings_11}
A.~A. Hagberg, D.~A. Schult, and P.~J. Swart.
\newblock Exploring network structure, dynamics, and function using networkx.
\newblock In Ga\"el Varoquaux, Travis Vaught, and Jarrod Millman, editors,
  \emph{Proceedings of the 7th Python in Science Conference}, pages 11 -- 15,
  Pasadena, CA USA, 2008.

\bibitem[Harris et~al.(2020)Harris, Millman, van~der Walt, Gommers, Virtanen,
  Cournapeau, Wieser, Taylor, Berg, Smith, Kern, Picus, Hoyer, van Kerkwijk,
  Brett, Haldane, del R{\'{i}}o, Wiebe, Peterson, G{\'{e}}rard-Marchant,
  Sheppard, Reddy, Weckesser, Abbasi, Gohlke, and Oliphant]{harris2020array}
C.~R. Harris, K.~J. Millman, S.~J. van~der Walt, R.~Gommers, P.~Virtanen,
  D.~Cournapeau, E.~Wieser, J.~Taylor, S.~Berg, N.~J. Smith, R.~Kern, M.~Picus,
  S.~Hoyer, M.~H. van Kerkwijk, M.~Brett, A.~Haldane, J.~Fern{\'{a}}ndez del
  R{\'{i}}o, M.~Wiebe, P.~Peterson, P.~G{\'{e}}rard-Marchant, K.~Sheppard,
  T.~Reddy, W.~Weckesser, H.~Abbasi, C.~Gohlke, and T.~E. Oliphant.
\newblock Array programming with {NumPy}.
\newblock \emph{Nature}, 585\penalty0 (7825):\penalty0 357--362, September
  2020.
\newblock \doi{10.1038/s41586-020-2649-2}.
\newblock URL \url{https://doi.org/10.1038/s41586-020-2649-2}.

\bibitem[Hoyer et~al.(2009)Hoyer, Janzing, Mooij, Peters, and
  Sch\"olkopf]{Hoyer}
P.~Hoyer, D.~Janzing, J.~Mooij, J.~Peters, and B~Sch\"olkopf.
\newblock Nonlinear causal discovery with additive noise models.
\newblock In D.~Koller, D.~Schuurmans, Y.~Bengio, and L.~Bottou, editors,
  \emph{Proceedings of the conference Neural Information Processing Systems
  (NIPS) 2008}, Vancouver, Canada, 2009. MIT Press.

\bibitem[Imbens and Rubin(2015)]{Imbens2015}
G.~W. Imbens and D.~B. Rubin.
\newblock \emph{Causal Inference for Statistics, Social, and Biomedical
  Sciences: An Introduction}.
\newblock Cambridge University Press, 2015.

\bibitem[Janzing et~al.(2013)Janzing, Balduzzi, Grosse-Wentrup, and
  Sch\"olkopf]{causalstrength}
D.~Janzing, D.~Balduzzi, M.~Grosse-Wentrup, and B.~Sch\"olkopf.
\newblock Quantifying causal influences.
\newblock \emph{Annals of Statistics}, 41\penalty0 (5):\penalty0 2324--2358,
  2013.

\bibitem[Janzing et~al.(2020)Janzing, Minorics, and
  Bl{\"o}baum]{janzing2020feature}
D.~Janzing, L.~Minorics, and P.~Bl{\"o}baum.
\newblock Feature relevance quantification in explainable {AI}: A causal
  problem.
\newblock In S.~Chiappa and R.~Calandra, editors, \emph{Proceedings of the
  Twenty Third International Conference on Artificial Intelligence and
  Statistics}, volume 108 of \emph{Proceedings of Machine Learning Research},
  pages 2907--2916, Online, 26--28 Aug 2020. PMLR.

\bibitem[Janzing et~al.(2024)Janzing, Bl\"{o}baum, A~Mastakouri, M~Faller,
  Minorics, and Budhathoki]{pmlr-v238-janzing24a}
D.~Janzing, P.~Bl\"{o}baum, A.~A~Mastakouri, P.~M~Faller, L.~Minorics, and
  K.~Budhathoki.
\newblock Quantifying intrinsic causal contributions via structure preserving
  interventions.
\newblock In Sanjoy Dasgupta, Stephan Mandt, and Yingzhen Li, editors,
  \emph{Proceedings of The 27th International Conference on Artificial
  Intelligence and Statistics}, volume 238 of \emph{Proceedings of Machine
  Learning Research}, pages 2188--2196. PMLR, 02--04 May 2024.
\newblock URL \url{https://proceedings.mlr.press/v238/janzing24a.html}.

\bibitem[Kalainathan and Goudet(2019)]{cdt}
D.~Kalainathan and O.~Goudet.
\newblock Causal discovery toolbox: Uncover causal relationships in {Python},
  2019.
\newblock URL \url{https://arxiv.org/abs/1903.02278}.

\bibitem[Lundberg and Lee(2017)]{Lundberg2017}
S.~Lundberg and S.~Lee.
\newblock A unified approach to interpreting model predictions.
\newblock In I.~Guyon, U.~V. Luxburg, S.~Bengio, H.~Wallach, R.~Fergus,
  S.~Vishwanathan, and R.~Garnett, editors, \emph{Advances in Neural
  Information Processing Systems 30}, pages 4765--4774. Curran Associates,
  Inc., 2017.

\bibitem[{M}c{K}inney(2010)]{mckinney-proc-scipy-2010}
{W}. {M}c{K}inney.
\newblock {D}ata {S}tructures for {S}tatistical {C}omputing in {P}ython.
\newblock In {S}t\'efan van~der {W}alt and {J}arrod {M}illman, editors,
  \emph{{P}roceedings of the 9th {P}ython in {S}cience {C}onference}, pages 56
  -- 61, 2010.
\newblock \doi{10.25080/Majora-92bf1922-00a}.

\bibitem[Miller et~al.(2020)Miller, Hsu, Troutman, Perdomo, Zrnic, Liu, Sun,
  Schmidt, and Hardt]{miller2020whynot}
J.~Miller, C.~Hsu, J.~Troutman, J.~Perdomo, T.~Zrnic, L.~Liu, Y.~Sun,
  L.~Schmidt, and M.~Hardt.
\newblock {WhyNot}, 2020.
\newblock URL \url{https://doi.org/10.5281/zenodo.3875775}.

\bibitem[Mothilal et~al.(2020)Mothilal, Sharma, and Tan]{mothilal2020dice}
R.~K Mothilal, A.~Sharma, and C.~Tan.
\newblock Explaining machine learning classifiers through diverse
  counterfactual explanations.
\newblock In \emph{Proceedings of the 2020 Conference on Fairness,
  Accountability, and Transparency}, pages 607--617, 2020.

\bibitem[Pearl(2009)]{Pearl2009}
J.~Pearl.
\newblock \emph{Causality: Models, Reasoning, and Inference}.
\newblock Cambridge University Press, New York, NY, 2nd edition, 2009.

\bibitem[Pedregosa et~al.(2011)Pedregosa, Varoquaux, Gramfort, Michel, Thirion,
  Grisel, Blondel, Prettenhofer, Weiss, Dubourg, Vanderplas, Passos,
  Cournapeau, Brucher, Perrot, and Duchesnay]{scikit-learn}
F.~Pedregosa, G.~Varoquaux, A.~Gramfort, V.~Michel, B.~Thirion, O.~Grisel,
  M.~Blondel, P.~Prettenhofer, R.~Weiss, V.~Dubourg, J.~Vanderplas, A.~Passos,
  D.~Cournapeau, M.~Brucher, M.~Perrot, and E.~Duchesnay.
\newblock Scikit-learn: Machine learning in {P}ython.
\newblock \emph{Journal of Machine Learning Research}, 12:\penalty0 2825--2830,
  2011.

\bibitem[Peters et~al.(2017)Peters, Janzing, and Sch\"olkopf]{causality_book}
J.~Peters, D.~Janzing, and B.~Sch\"olkopf.
\newblock \emph{Elements of Causal Inference -- Foundations and Learning
  Algorithms}.
\newblock MIT Press, 2017.

\bibitem[Ramsey et~al.()Ramsey, Zhang, Glymour, Romero, Huang, Ebert-Uphoff,
  Samarasinghe, Barnes, and Glymour]{ramsey2018tetrad}
J.~D Ramsey, K.~Zhang, M~Glymour, R.~S. Romero, B.~Huang, I.~Ebert-Uphoff,
  S.~Samarasinghe, E.~A Barnes, and C.~Glymour.
\newblock Tetrad—a toolbox for causal discovery.

\bibitem[Rubin(1974)]{Rubin1974}
D.~B. Rubin.
\newblock Estimating causal effects of treatments in randomized and
  nonrandomized studies.
\newblock \emph{Journal of Educational Psychology}, 66:\penalty0 688--701,
  1974.

\bibitem[Sharma and Kiciman(2020)]{dowhy}
A.~Sharma and E.~Kiciman.
\newblock {DoWhy}: An end-to-end library for causal inference, 2020.
\newblock URL \url{https://arxiv.org/abs/2011.04216}.

\bibitem[Virtanen et~al.(2020)Virtanen, Gommers, Oliphant, Haberland, Reddy,
  Cournapeau, Burovski, Peterson, Weckesser, Bright, {van der Walt}, Brett,
  Wilson, Millman, Mayorov, Nelson, Jones, Kern, Larson, Carey, Polat, Feng,
  Moore, {VanderPlas}, Laxalde, Perktold, Cimrman, Henriksen, Quintero, Harris,
  Archibald, Ribeiro, Pedregosa, {van Mulbregt}, and {SciPy 1.0
  Contributors}]{2020SciPy-NMeth}
P.~Virtanen, R.~Gommers, T.~E. Oliphant, M.~Haberland, T.~Reddy, D.~Cournapeau,
  E.~Burovski, P.~Peterson, W.~Weckesser, J.~Bright, S.~J. {van der Walt},
  M.~Brett, J.~Wilson, K.~J. Millman, N.~Mayorov, A.~R.~J. Nelson, E.~Jones,
  R.~Kern, E.~Larson, C~J Carey, {\.I}.~Polat, Y.~Feng, E.~W. Moore,
  J.~{VanderPlas}, D.~Laxalde, J.~Perktold, R.~Cimrman, I.~Henriksen, E.~A.
  Quintero, C.~R. Harris, A.~M. Archibald, A.~H. Ribeiro, F.~Pedregosa, P.~{van
  Mulbregt}, and {SciPy 1.0 Contributors}.
\newblock {{SciPy} 1.0: Fundamental Algorithms for Scientific Computing in
  Python}.
\newblock \emph{Nature Methods}, 17:\penalty0 261--272, 2020.
\newblock \doi{10.1038/s41592-019-0686-2}.

\bibitem[Zhang and Hyv\"arinen(2009)]{Zhang_UAI}
K.~Zhang and A.~Hyv\"arinen.
\newblock On the identifiability of the post-nonlinear causal model.
\newblock In \emph{Proceedings of the 25th Conference on Uncertainty in
  Artificial Intelligence}, Montreal, Canada, 2009.

\end{thebibliography}
\end{document}